\documentclass[12pt]{iopart}

\usepackage{graphics}
\usepackage{color}
\usepackage{graphicx}
\usepackage{dcolumn}
\usepackage{bm}
\usepackage{epsfig}
\usepackage{float}
\begin{document}


\title{Comparative study of DFT+$U$ functionals for non-collinear magnetism}

\author{Siheon Ryee}
\address{Department of Physics, Korea Advanced Institute of Science and Technology (KAIST), Daejeon 34141, Korea}
\author{Myung Joon Han}
\ead{mj.han@kaist.ac.kr}
\address{Department of Physics, Korea Advanced Institute of Science and Technology (KAIST), Daejeon 34141, Korea}
\date{\today}

\begin{abstract}
We performed comparative analysis for DFT+$U$ functionals to better understand their applicability to non-collinear magnetism. Taking LiNiPO$_4$ and Sr$_2$IrO$_4$ as examples, we investigated the results out of two formalisms based on charge-only density and spin density functional plus $U$ calculations. Our results show that the ground state spin order in terms of tilting angle is strongly dependent on Hund $J$. In particular, the opposite behavior of canting angles as a function of $J$ is found for LiNiPO$_4$. The dependence on the other physical parameters such as Hubbard $U$ and Slater parameterization $F^4/F^2$ is investigated. We also discuss the formal aspects of these functional dependences as well as parameter dependences. The current study provides 
useful information and important intuition for the first-principles calculation of non-collinear magnetic materials.
\end{abstract}

\maketitle


\section{Introduction} \label{intro} 

DFT+$U$ (density functional theory + $U$) method  \cite{Anisimov_91,Liechtenstein,LDAU_review} is a powerful tool to describe the correlated electron materials. Due to its computation efficiency and the straightforwardness of physical interpretation, DFT+$U$ has become one of the most widely-used schemes to study complex magnetic phases in solids. On the other hand, an obvious drawback of DFT+$U$ as a first-principles-based method is that the result is dependent on the interaction parameters such as Hubbard $U$ and Hund $J$. The choice of double counting energy and density-functional type poses another important issue for DFT+$U$ and related techniques such as DFT+DMFT (DFT + dynamical mean-field theory).

A series of recent studies \cite{JChen,Park,Chen,SR} have made notable progress in this regard. Careful investigations of the $J$ dependence and the exchange-correlation (XC) functional dependence of widely-used DFT+$U$ formulations show that adopting spin-density XC functional, such as LSDA (local spin-density approximation) and SGGA (spin-polarized generalized gradient approximation), can cause undesirable results. For clarity, we hereafter denote this type of formulation by `SDFT+$U$' (spin-density functional theory + $U$). The unphysical feature within SDFT+$U$ originates from the intrinsic exchange energy within spin-density XC functionals. In order to avoid such an artifact, the use of charge density (i.e., spin-unpolarized) XC functional (denoted by `CDFT+$U$' to include both charge-only LDA+$U$ and GGA+$U$ simultaneously) is desirable \cite{JChen,Park,Chen,SR}. Although many detailed features have been formally analyzed recently \cite{SR}, one important question still remains: Since all of these studies only dealt with collinear spin density, the case for non-collinear spin orders is basically unexplored to the best of our knowledge. While a previous study \cite{Spaldin} provides useful information regarding the non-collinear magnetism within SDFT+$U$, more analysis at a deeper level and its comparison with CDFT+$U$ results is still elusive. Not only because the non-collinear spin order is quite often realized in real materials \cite{Nagaosa,Anjan}, but it also affects the other physical property such as electrical polarization in magneto-electric materials \cite{Fiebig,Katsura,Birol,Pyatakov}, a systematic investigation is strongly requested.

In this paper, we performed a systematic comparative study of DFT+$U$ formalisms for the case of non-collinear magnetic orders. With LiNiPO$_4$ (LNPO) and Sr$_2$IrO$_4$ (SIO) as our examples, we investigated the magnetic ground state property predicted by SDFT+$U$ and CDFT+$U$, and its dependence on physical as well as numerical parameters. We show that the two formalisms give the opposite trend of spin canting angle as a function of $J$ while the $U$ dependence is weak in LNPO. The difference is mainly attributed to the different treatment of spin-off-diagonal potential terms. 
For the case of SIO, the canting angle exhibits a stronger dependence on the choice of $U$ while it exhibits the same increasing trend as a function of $J$ in both formalisms. The dependence on the Slater parameterization and the local projector is also investigated.
Our results clearly show that the current DFT+$U$ schemes have limitations in describing non-collinear magnetism, and further methodological development is requested. As a step toward this direction, we present and discuss our analysis of the two functionals in non-collinear spin space.

\begin{figure*} [!htbp]
	\begin{center}	
	\includegraphics[width=0.9\textwidth, angle=0]{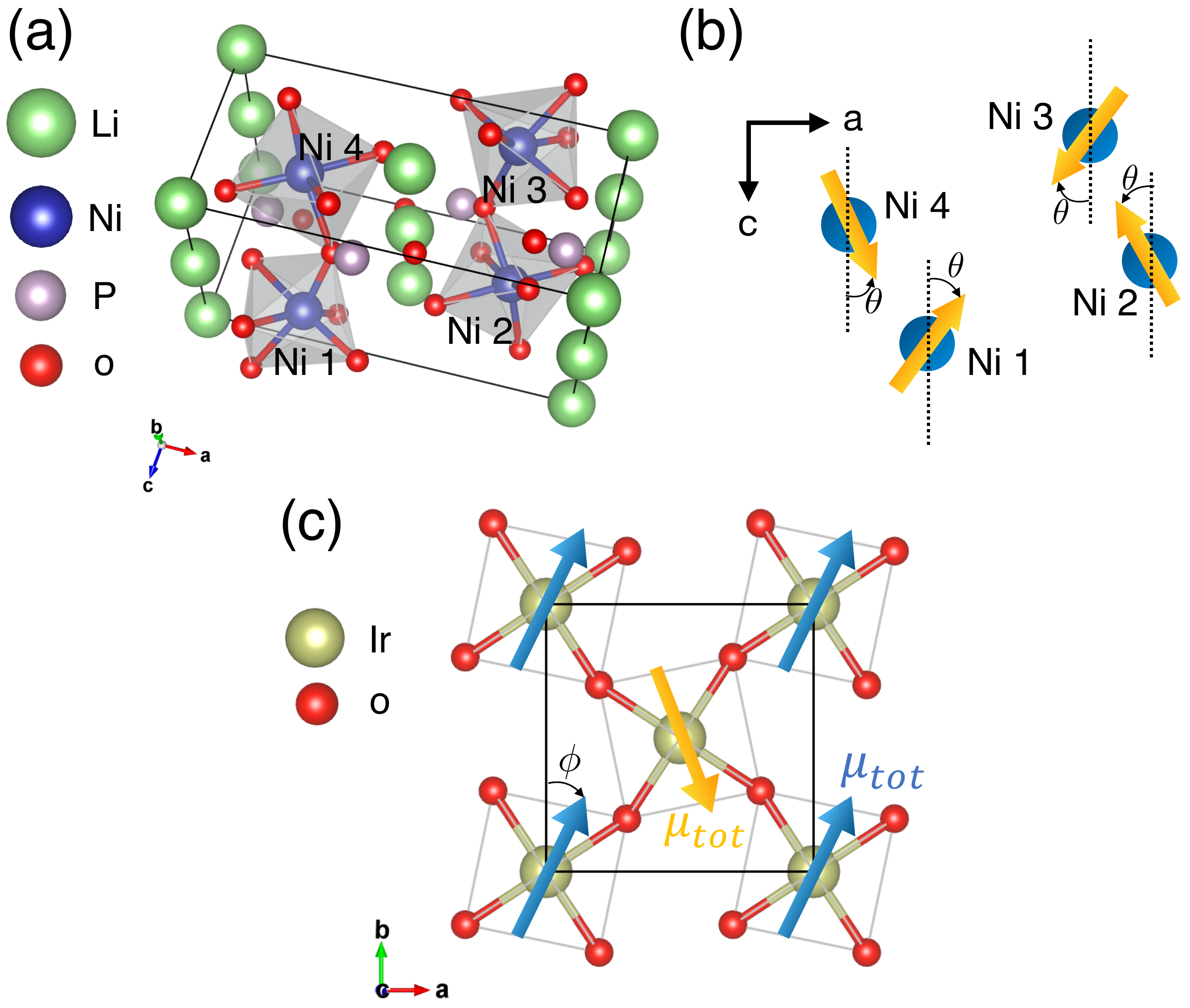}
	\caption{(a) Crystal structure of LNPO. For the positions of Ni-sites, we follow Ref.~\cite{Jensen}; Ni~1~$=(0.275, 0.25, 0.98)$, Ni~2~$=(0.775, 0.25, 0.52)$, Ni~3~$=(0.725, 0.75, 0.02)$, and Ni~4~$=(0.225, 0.75, 0.48)$. (b) A schematic view of non-collinear spin alignment in LNPO. Spins lie within $ac$-plane. The direction of angle $\theta$ here represents the positive value. (c) Canted-AFM alignment of $\mu_\mathrm{tot}$ (the sum of spin and orbital moment at each Ir site) in SIO. Moments lie within IrO$_2$ planes and oxygen octahedra are rotated by $11.5^\circ$ with respect to $c$-axis \cite{Crawford}. }
	\label{fig1}
	\end{center}
\end{figure*}

\section{Computation Method}

All calculations were performed using rotationally invariant version of DFT+$U$ method \cite{Liechtenstein,LDAU_review} as implemented in `OpenMX' software package \cite{SR, openmx}. Our formalism is based on the non-orthogonal LCPAO (linear combination of localized pseudoatomic orbitals) \cite{LCPAO1,LCPAO2,LCPAO3}. The effect of SOC (spin-orbit coupling) was treated within the fully relativistic pseudopotentials \cite{openmx,MBK}. Experimental crystal structure was used for LNPO ($Pnma$) and SIO ($I4_1/acd$) with lattice parameters of $a=10.02$, $b=5.83$, $c=4.66$ {\AA} \cite{lnpo_structure},  and $a=b=5.48$, $c=25.80$ {\AA} \cite{Crawford}, respectively (see Fig.~\ref{fig1}). We adopted 3 $\times$ 5 $\times$ 5 (7 $\times$ 7 $\times$ 3) $\mathbf{k}$-points in the first Brillouin zone of LNPO (SIO) and the energy cutoff of 500 Ry in real space grid for numerical integrations. An energy criterion of $10^{-9}$ Hartree was used for the self-consistency. The radial cutoff of PAOs were set to 8.0, 6.0, 7.0, 10.0, 7.0 and 5.0 a.u. for Li, Ni, P, Sr, Ir, and O, respectively. For XC functional, L(S)DA \cite{CA} parameterized by Perdew and Zunger \cite{CA-PZ} was used. Unless otherwise specified, the `dual' projector \cite{MJH} was used for the on-site density matrix (DM), and the $U$ value was set to 5 eV for LNPO and 2 eV for SIO following the previous DFT+$U$ calculations \cite{Spaldin,Yamauchi,BJKim,Jin}. 

The total energy correction by DFT+$U$, $E^U$, can be expressed as:
\begin{eqnarray} 
E^U = E^{\mathrm{int}} - E^\mathrm{dc},
\end{eqnarray}
where $E^\mathrm{int}$ represents the Hubbard-type on-site interaction energy \cite{Liechtenstein}.
For the double counting term ($E^\mathrm{dc}$), we adopted the so-called FLL (fully localized limit) functional. For the case of SDFT+$U$ it can be written as \cite{Liechtenstein,SR,Czyzyk}
\begin{eqnarray} \label{sFLL}
E_{\mathrm{SDFT}+U}^{\mathrm{dc,FLL}} = \frac{1}{2}UN(N-1) - \frac{1}{2}JN\bigg(\frac{N}{2}-1\bigg) -\frac{1}{4}J{\vec{\mathrm{M}} \cdot \vec{\mathrm{M}}}.
\end{eqnarray}
It is worth to be noted that the widely-used Dudarev's approach \cite{Dudarev} is equivalent to the $J=0$ version of SDFT+$U$ employing FLL double counting \cite{Liechtenstein,SR, Himmentoglu}.
For CDFT+$U$, it reads \cite{SR,Anisimov_93,Solovyev_94} 
\begin{eqnarray} \label{cFLL}
E_{\mathrm {CDFT}+U}^{\mathrm{dc,FLL}} = \frac{1}{2}UN(N-1) - \frac{1}{2}JN\bigg(\frac{N}{2}-1\bigg),
\end{eqnarray}
where the $d$-electron occupation is $N=\mathrm{Tr}[\mathbf{n}]$, and $n^{\sigma\sigma'}_{m_1m_2}$ are the elements of on-site DM, $\mathbf{n}$, for given orbitals $\{m_i\}$ and spins $\sigma,\sigma'$ ($\sigma,\sigma' = \uparrow$ or $\downarrow$). Note that the correlated orbtials are responsible for the magnetization in CDFT+$U$, and the rest space is affected by hybridization with them. The spin moment is given by $\vec{\mathrm{M}} = \mathrm{Tr}[\vec{\sigma} \mathbf{n}]$ where $\vec{\sigma}$ is Pauli matrices. For clarity we use `sFLL' and `cFLL' in referring to `SDFT+$U$' and `CDFT+$U$' (expressed with the FLL double counting functional), respectively, throughout the rest of this paper. For more formulation details, see Ref.~\cite{SR} and references therein.

\section{Result and Discussion}
\subsection{LiNiPO$_4$}

The Li orthophosphates Li{\it M}PO$_4$ ({\it M} = Mn, Fe, Co, Ni) are magneto-electric materials \cite{Jensen,Kornev,Vaknin}. A strong magneto-electric effect is observed in antiferromagnetic (AF) phase of Li{\it M}PO$_4$. In the present study, we focus on the non-collinear magnetic ground state of LNPO. Neutron diffraction shows that `$C_zA_x$'-type ($C$-type AF order along $c$-axis and $A$-type along $a$-axis with small cantings) spin alignment is stabilized at low temperature \cite{Jensen} (see Fig.~\ref{fig1}(b)).

\begin{figure*} [!htbp]
	\begin{center}
	\includegraphics[width=0.9\textwidth, angle=0]{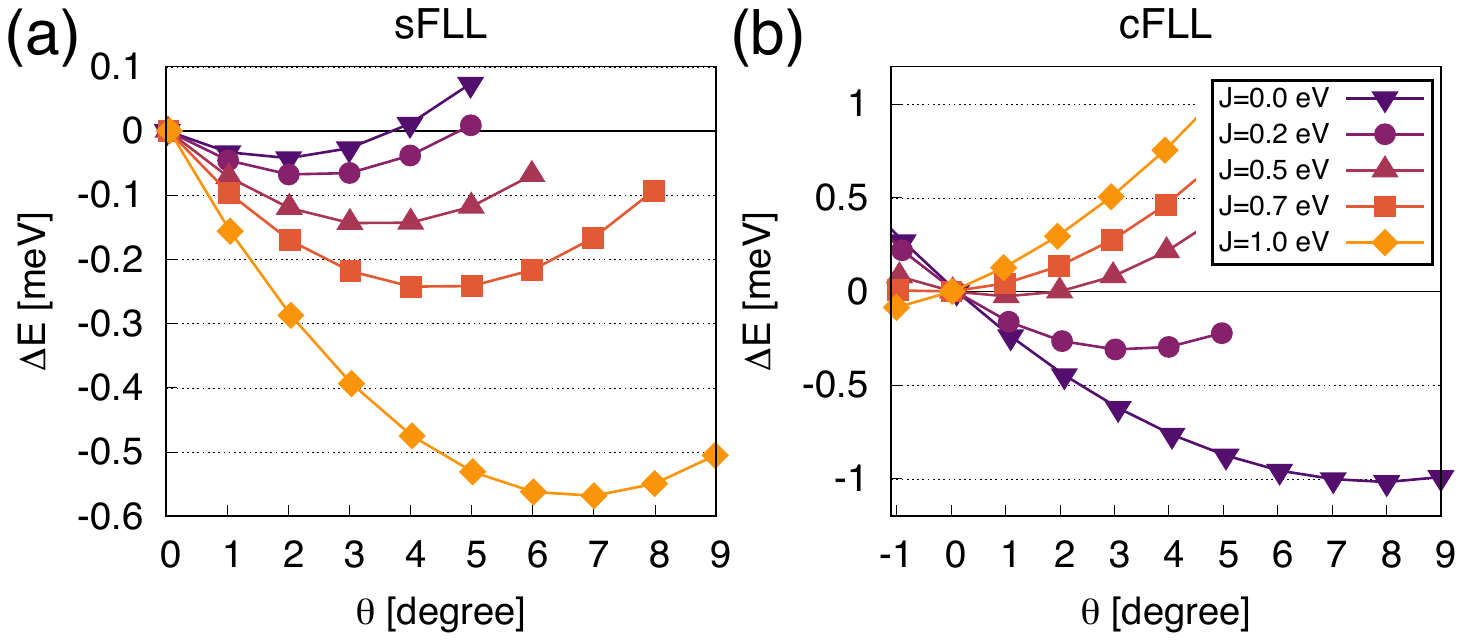}
	\caption{The total energies (per four formula units) as a function of spin canting angle $\theta$ and Hund $J$ calculated by (a) sFLL and (b) cFLL. Note that the spin with negative $\theta$ points to the opposite canting direction (see Fig.~\ref{fig1}(b)). The zero energy is set to the results of uncanted ($\theta$=0) solution for each $J$. }
	\label{fig2}
	\end{center}
\end{figure*}

To investigate the predictive capability of current DFT+$U$ methods for the non-collinear magnetic ground state, we performed the constrained spin-moment total energy calculations \cite{openmx,Kurz,Ma} in which the spins are restricted to lie within the $ac$-plane (as known from experiment) with an angle $\theta$ (as shown in Fig.~\ref{fig1}(b)). Note that no constraint is imposed on orbital moment. The results are summarized in Fig.~\ref{fig2}  where the zero energy points correspond to the uncanted (i.e., collinear) spin order. Since the stable spin configuration is sensitive to $J$ as reported by Bousquet and Spaldin \cite{Spaldin}, five different $J$ values in the range of $0\leq J\leq 1$ eV have been considered.

First of all, we note the different $J$ dependence found in between sFLL (Fig.~\ref{fig2}(a)) and cFLL (Fig.~\ref{fig2}(b)). In sFLL, the larger canting angle is favored for the larger $J$ with the greater stabilization energy: At $J=0$ eV, the canting angle is $\theta = 1.88^\circ$ and it gradually increases to $\theta =6.76^\circ$ at $J=1$ eV. In comparison to the experimental value of $\theta \simeq 7.8^\circ$ \cite{Jensen}, the best agreement is expected at $J>1$ eV which might be too large to be realistic. In the entire range of $J$ we considered, the moment size change is negligible; $\mu_s\simeq 1.8$ $\mu_{B}/$Ni.

\begin{figure*} [!htbp]
	\begin{center}
	\includegraphics[width=1.0\textwidth, angle=0]{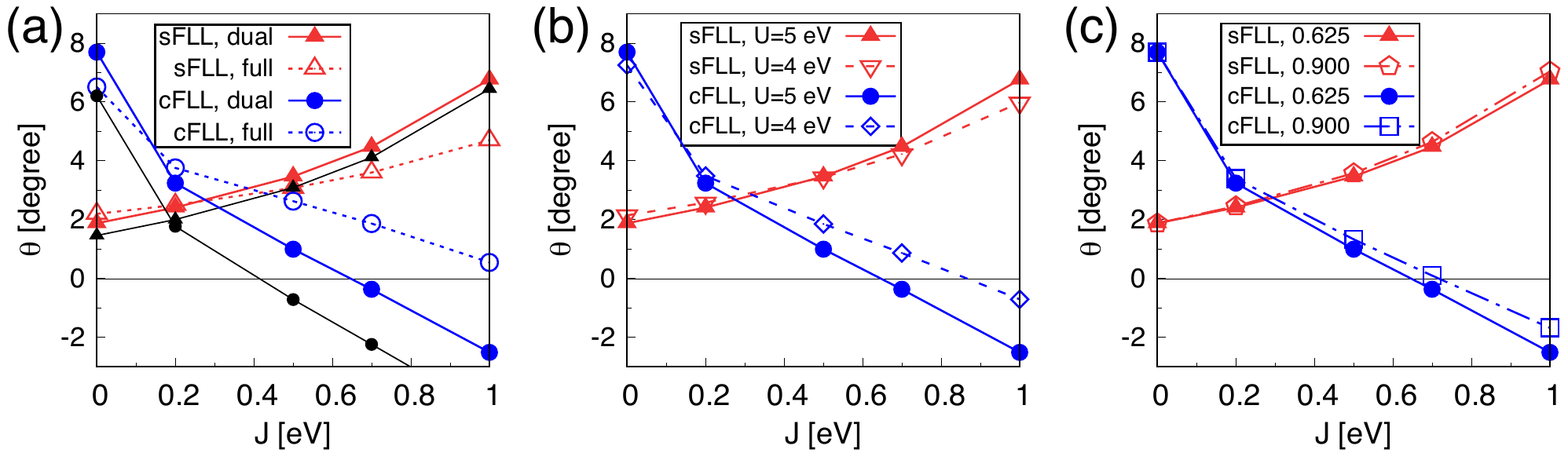}
	\caption{(a) The most stable non-collinear spin canting angle $\theta$ as a function of $J$. The results obtained by different functionals (sFLL (triangles) and cFLL (circles)) and by different projectors (`dual (filled symbols; solid lines)' and `full (empty symbols; dotted lines)') are compared. The black solid lines represent the total moment (the sum of spin and orbital moments) for cFLL (circles) and sFLL (triangles) obtained from dual projector. (b) The effect of $U$ value change; $U=5$ (solid lines) and 4 eV (dashed lines). (c) The effect of $F^4/F^2$ ratio change; $F^4/F^2$=0.625 (solid lines) and $F^4/F^2$=0.900 (dash-dotted lines).}
	\label{fig3}
	\end{center}
\end{figure*}

In cFLL, on the other hand, the canting angle is reduced as $J$ increases, and it eventually goes to $\theta=0$ at $J \simeq 0.7$ eV. For $J > 0.7$ eV, $\theta$ becomes negative (e.g., $\theta \simeq -2.9^\circ$ at $J=1$ eV), which means the canting angles are opposite along $a$ axis. Contrary to sFLL, the size of spin moment is gradually increased as $J$ increases as discussed in Ref.~\cite{SR}; from $\mu_s =1.57$ $\mu_B/$Ni at $J=0$ eV to $\mu_s =1.73$ $\mu_B/$Ni at $J=1$ eV. The experimental value of $\theta \simeq 7.8^\circ$ is achieved at $J\simeq 0$ eV which is also unrealistic.

The contribution from orbital moment is notably smaller than the spin (see black lines in Fig.~\ref{fig3}(a)); $\mu_{o}\sim 0.26$ $\mu_B/$Ni in the entire region of $J$ within sFLL while it increases from 0.50 $\mu_B/$Ni ($J=0$ eV) to 0.65 $\mu_B/$Ni  ($J=1$ eV) within cFLL.

The $U$ dependence of canting angle is less pronounced as reported in Ref.~\cite{Spaldin}. Our results of $U=4$ and $5$ eV are presented in Fig.~\ref{fig3}(b). The size of spin moment for $U=4$ eV is only slightly reduced from the $U=5$ eV result by $\sim 0.01 \mu_B/$Ni.

We also checked the effect of using different Slater parameterization in generating Coulomb interaction tensor. This parameterization has not attracted much attention in DFT+$U$ methodology. As studied in Ref.~\cite{Vaugier, Bultmark}, however, due to the screening in solids, $F^4/F^2$ can be enhanced and thus be different from the conventional value of $F^4/F^2=0.625$ which corresponds to the atomic environment. To further elucidate this point within the context of non-collinear formalism, we performed the calculations with $F^4/F^2=0.9$ and 
the result is compared with that of $F^4/F^2=$0.625 in Fig.~\ref{fig3}(c). 
It is found that the canting-angle change caused by different parameter choice is not significant; At $J=1$ eV, $\Delta\theta\sim +0.8^\circ$ for cFLL and $\sim +0.3^\circ$ for sFLL. The moment size change is also quite small, $\sim 0.01\mu_B$, in both cFLL and sFLL.

As a summary of current subsection, we note that, with the current DFT+$U$ formalisms, it is difficult to determine or predict the non-collinear spin angle. The $J$ dependence is noteworthy.
Considering the cRPA (constrained random phase approximation) estimation of $J=0.7$ -- $0.9$ eV for (isovalent) NiO \cite{Sakuma,Panda}, sFLL and cFLL gives $\theta \simeq +4^\circ$ -- $+6^\circ$ and $\simeq 0^\circ$ -- $-2^\circ$, respectively. Further, the origin of opposite behavior of canting angles is far from clear. This issue together with the physical meaning of atomic Hund interaction certainly deserves further investigations.

\subsection{Sr$_2$IrO$_4$}

The second example is SIO which is known as a `relativistic Mott' insulator \cite{BJKim,BJKim2} exhibiting a canted AF order with a canting angle $\phi \sim 12-13^\circ$ within a IrO$_2$ plane (see Fig.~\ref{fig1}(c)) \cite{Ye,Boseggia}. 
Its low energy behavior can be described by so-called $J_\mathrm{eff}=1/2$ Kramers doublet; namely, $|\tilde{\uparrow}\rangle = \sin\theta_t ~|0,\uparrow\rangle - \cos\theta_t ~|+1,\downarrow\rangle$ and $|\tilde{\downarrow}\rangle = \sin\theta_t ~|0,\downarrow\rangle - \cos\theta_t ~|-1,\uparrow\rangle$ where $\theta_t$ is parameterized via $\tan(2\theta_t)=2\sqrt{2}\lambda/(\lambda-2\Delta)$ \cite{Jackeli}. Here, $\lambda$ and $\Delta$ refers to the SOC strength and the tetragonal splitting within $t_{2g}$ manifold, respectively, and $|0\rangle=|d_{xy}\rangle$ and $|\pm1\rangle=-\frac{1}{\sqrt{2}}(i|d_{xz}\rangle \pm |d_{yz}\rangle)$. 
The cooperation of isotropic superexchange and Dzyaloshinskii-Moriya interaction leads to the canted AF state carrying net ferromagnetic moment of $M \sim 0.1 - 0.2$ $\mu_{B}$/Ir \cite{BJKim2,Ye} along $a$-direction (Fig.~\ref{fig1}(c)).

\begin{figure*} [!htbp]
	\begin{center}
	\includegraphics[width=1.0\textwidth, angle=0]{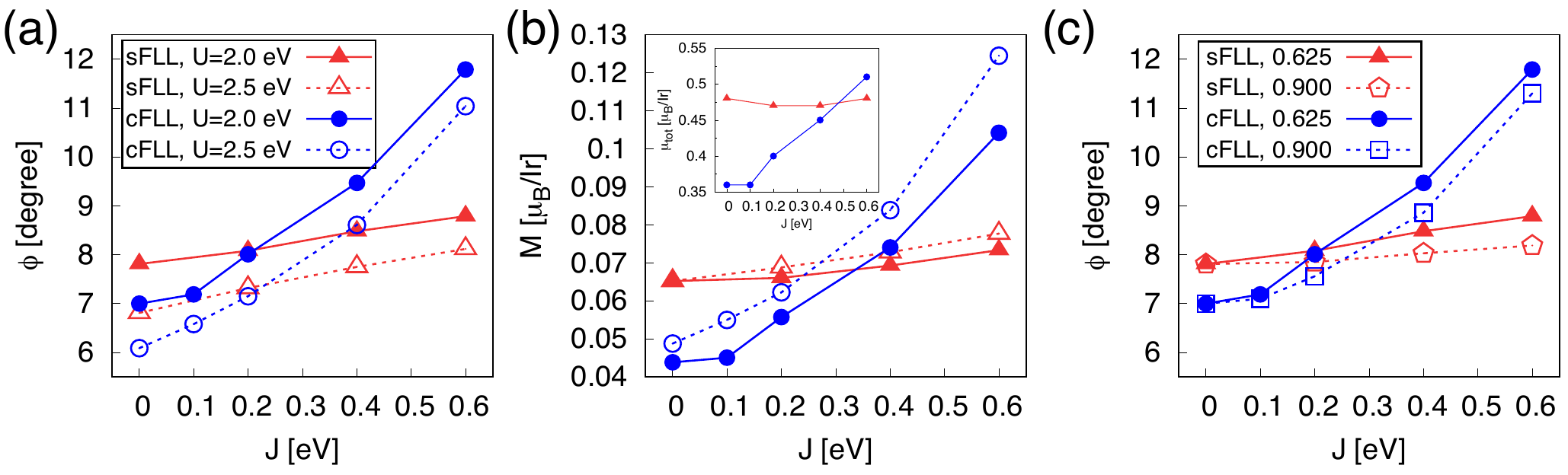}
	\caption{(a) The calculated canting angle $\phi$ of total moment ($\mu_\mathrm{tot}$; see Fig.~\ref{fig1}(c)) as a function of $J$. Two different $U$ values ( i.e., $U$=2 (solid lines) and 2.5 eV (dashed lines)) and functionals (i.e., sFLL (triangles) and cFLL (circles)) have been considered. (b) The calculated net moment ($M$) as a function of $J$ depending on the functionals and $U$ values (same notations with (a)). The inset shows the size of $\mu_\mathrm{tot}$ as a function of $J$ at $U=2$ eV. (c) The calculated results of $\phi$ with two different $F^4/F^2$ ratio; $F^4/F^2$=0.625 (solid lines) and $F^4/F^2$=0.900 (dash-dotted lines).}
	\label{fig4}
	\end{center}
\end{figure*}

The calculation results of canting angle are summarized in Fig.~\ref{fig4}(a). 
Here we present the angle $\phi$ of total moment $\mu_\mathrm{tot}$ (the sum of spin and orbital moment at each Ir site) as a function of $J$ (see Fig.~\ref{fig1}(c)). On the contrary to the case of LNPO, sFLL and cFLL exhibit the same increasing trend of canting angle as $J$ increases while cFLL gives a better agreement with the experimental value of $\phi\sim 12-13^\circ$ \cite{Ye,Boseggia} at realistic $J$ (see Fig.~\ref{fig4}(a)). Note that the canting angle is sensitive to the structural distortion which is fixed to the experimental value in our calculations \cite{Liu}.

The choice of $U$ and $J$ can significantly affect the net moment size as shown in Fig.~\ref{fig4}(b). The cFLL result at $J=0.6$ eV is in a good agreement with experimental value of $M \sim 0.1 - 0.2$ $\mu_{B}$/Ir \cite{BJKim2,Ye}. The net moment enhancement as a function of $J$ is originated from both the enhanced total moment $\mu_\mathrm{tot}$ and the enlarged canting angle. See the inset of Fig.~\ref{fig4}(b) which shows that, for $U=2$ eV, the total moment $\mu_\mathrm{tot}$ is increased from $\mu_\mathrm{tot}=0.35$ ($J=0$ eV) to 0.5 $\mu_{B}$/Ir ($J=0.6$ eV) within cFLL while $\mu_\mathrm{tot} \simeq 0.47$ $\mu_{B}$/Ir within sFLL in the entire range of $J$ .

The $U$ dependence of canting angle is also presented in Fig.~\ref{fig4}(a). The larger value of $U$ by 0.5 eV results in the reduced canting angle by $\sim 1^\circ$  in the entire $J$ range we considered for both sFLL and cFLL. The $U$ dependence of moment size can be found in Fig.~\ref{fig4}(b). For $\mu_\mathrm{tot}$ (not shown), it is increased by $\sim 0.12\mu_{B}$/Ir ($\sim 0.07$ $\mu_{B}$/Ir) within cFLL (sFLL) when we use $U=2.5$ eV instead of 2.0 eV. For the case of sFLL, this is a comparable size of change caused by varying $J$.

The results of $F^4/F^2$ dependence are presented in Fig.~\ref{fig4}(c). In both formalisms of sFLL and cFLL, it is found that the canting angle is reduced by $\sim 0.5^\circ$ when using $F^4/F^2$=0.9 at $J=0.6$ eV. It is worth to be noted that increasing $F^4/F^2$ ratio leads to the change of off-diagonal elements in the Coulomb interaction matrix by about $\pm 0.01J$ (see Sec.~\ref{analysis}). Considering that this is a small change in terms of interaction strengths, the canting angle is sensitive to the parameterization of Slater integrals especially in sFLL for which $J$ dependence is much weaker.

\subsection{Further analysis and discussion} \label{analysis}

Our results of two material examples clearly show that standard DFT+$U$ formalisms can give quite different solutions. It also implies that there is no {\it a priori} rule in choosing a functional for the description of non-collinear magnetism. For LNPO, sFLL seems to give a better agreement with experiment whereas cFLL is the case for SIO. The effect of $U$ on the canting angle is material dependent in both formalisms; it is basically negligible in LNPO but not in SIO.

Thus our result raises a question regarding the predictive power of current DFT+$U$ methodology for non-collinear magnetism. Since there is no good intuitive reason available for these behaviors, one needs to be careful when there is no guide from outside, e.g., from experiments. Further, as can be noticed in the cases of LaMnO$_3$ \cite{Mellan}, Li orthophosphates, and difluorites \cite{Spaldin}, reproducing the correct electronic and spin structures does not always guarantee the physically relevant $U$ and $J$ range. In particular, within SDFT+$U$ scheme, unphysically large $J$ values can be required \cite{Spaldin, Mellan}. It can also be true in the non-collinear case (as in the collinear case) that some redundant contributions reside in LSDA/SGGA on top of atomic on-site interactions \cite{JChen,Park,Chen,SR}. However, the usual sense of `interactions' does not seem directly responsible for non-collinearity, and therefore our intuition of choosing those parameters is not quite helpful any longer in this situation.

At this moment, it is quite challenging to fully understand the physical meaning of the observed differences in between sFLL and cFLL, and to construct the revised DFT+$U$ functional compatible with non-collinear spin density. As a step towards this direction, we further analyze below the origin of this behavior within the currently available standard formulation. In the non-collinear spin DFT, the XC potential is given by \cite{LSDA,MacDonald,Kubler}
\begin{eqnarray} \label{XC}
\mathbf{V}_\mathrm{XC} &= 
\left(
\begin{array}{cc} 
V^{\uparrow \uparrow}_\mathrm{XC} & V^{\uparrow \downarrow}_\mathrm{XC} \nonumber \\ V^{\downarrow \uparrow}_\mathrm{XC} & V^{\downarrow \downarrow}_\mathrm{XC} 
\end{array} 
\right) \nonumber \\
&=
\left(
\begin{array}{cc} 
V^0_\mathrm{XC}+\Delta V_\mathrm{XC}\cos\beta & \Delta V_\mathrm{XC}\exp(-i\alpha)\sin\beta \\ \Delta V_\mathrm{XC}\exp(i\alpha)\sin\beta  & V^0_\mathrm{XC}-\Delta V_\mathrm{XC}\cos\beta 
\end{array}
\right),
\end{eqnarray}
where $\alpha$ and $\beta$ are the azimuthal and polar Euler angles, respectively. $V^0_\mathrm{XC} = (V^\uparrow_\mathrm{XC}+V^\downarrow_\mathrm{XC})/2$ and $\Delta V_\mathrm{XC} = (V^\uparrow_\mathrm{XC}-V^\downarrow_\mathrm{XC})/2$ is the charge potential and the spin up/down potential difference, respectively, obtained from the diagonalization of non-collinear spin-density matrix:
\begin{eqnarray}
\mathbf{n}= 
\left(
\begin{array}{cc} 
n^{\uparrow \uparrow} & n^{\uparrow \downarrow} \\ n^{\downarrow \uparrow} & n^{\downarrow \downarrow} 
\end{array}
\right).
\end{eqnarray}
It is obvious from Eq.~(\ref{XC}) that $\alpha=\beta=0$ is reduced back to the usual collinear spin-density functional.

Note that, in cFLL (i.e., with spin-unpolarized charge density in XC functional), $\Delta V_\mathrm{XC}=0$. As a result, the off-diagonal double counting potential should vanish for cFLL; $V_{m_1m_2}^{\mathrm{dc},\sigma \overline{\sigma}} = {\partial{E^{\mathrm{dc,FLL}}_{\mathrm{CDFT}+U}}}/{\partial n^{\sigma \overline{\sigma}}_{m_1m_2}}=0$ (where $\overline{\sigma}$ denotes the opposite spin to $\sigma$). For sFLL, on the other hand,  $V^{\mathrm{dc},\sigma \overline{\sigma}}_{m_1m_2}={\partial{E^{\mathrm{dc,FLL}}_{\mathrm{SDFT}+U}}}/{\partial n^{\sigma \overline{\sigma}}_{m_1m_2}} = -JN^{\overline{\sigma} \sigma}\delta_{m_1 m_2}$ (where $N^{\overline{\sigma} \sigma}=\mathrm{Tr}[\textbf{n}^{\overline{\sigma} \sigma}]$), which is not canceled out in general by $\Delta V_\mathrm{XC}$ term in $V^{\sigma \overline{\sigma}}_\mathrm{XC}$. As reported in Ref.~\cite{Spaldin}, enforcing $V^{\mathrm{dc},\sigma \overline{\sigma}}_{m_1m_2}=0$ within sFLL affects the spin canting. However, this treatment requires a careful physical interpretation because the contribution from spin-density XC functional should have been subtracted to avoid double counting. On the contrary, in cFLL, the off-diagonal spin potential and energy contain only $E^\mathrm{int}$-related terms due to $\Delta V_\mathrm{XC}=0$ (see Eq.~(\ref{pot})).

Since the off-diagonal potential is affected by $\Delta V_\mathrm{XC}$, it is useful to compare LSDA result with SGGA at $J=0$. SGGA has a stronger tendency toward the magnetic solution than LSDA, and the larger XC potential is expected in SGGA than LSDA \cite{Ryee}; $|\Delta V^\mathrm{SGGA}_\mathrm{XC}| > |\Delta V^\mathrm{LSDA}_\mathrm{XC}|$. It again affects the non-collinear spin potential through Eq.~(\ref{XC}). 
We found that the most stable canting angle of LNPO is $\theta =5.41^\circ$ and $\theta = 2.14^\circ$ within LSDA and SGGA (PBE parameterization \cite{PBE}), respectively, at $U=0$ eV. For the case of SIO,  the difference is small; $\phi =7.81^\circ$ and $7.94^\circ$ in LSDA and SGGA, respectively. Note that the enhanced $|\Delta V_\mathrm{XC}|$ gives the less (more) canted solution for LNPO (SIO). Thus, this feature is consistent with the result of cFLL (Fig.~\ref{fig2}(b) and Fig.~\ref{fig4}(b))) in which the spin-off-diagonal potential \cite{SR},
\begin{eqnarray} \label{pot}
V_{m_1m_2}^{U,\sigma \overline{\sigma}} &= - \sum_{m_3,m_4}\langle m_1,m_3|V_{ee}|m_4,m_2 \rangle n^{\overline{\sigma} \sigma}_{m_3m_4},
\end{eqnarray}
has the terms that are proportional to the elements of Coulomb interaction tensor, 
\begin{eqnarray} \label{Coulomb}
&\langle m_1,m_3|V_{ee}|m_4,m_2 \rangle  \nonumber \\
&= \sum_{\{m_i'\}}\Big[S_{m_1m_1'}S_{m_3m_3'} \Big\{\sum_{k=0}\alpha_k(m_1',m_3',m_2',m_4')F^k\Big\}
S^{-1}_{m_2'm_4}S^{-1}_{m_4'm_2} \Big].
\end{eqnarray}
Note that the magnitude of matrix elements increases with $J$ (see Eq.~(\ref{coulomb})). Here $F^k$ and $\alpha_k$ are Slater integrals and Racah-Wigner numbers, respectively, and $S$ is a transformation matrix from spherical harmonics to the pre-defined local basis \cite{Liechtenstein,Czyzyk,Vaugier} (see, e.g., Ref.~\cite{Vaugier} for more discussion on Coulomb interaction in transition metal oxides).

Another important issue is the on-site electron density. How to define the correlated subspace has been an important question in DFT+$U$ formalism \cite{MJH,full,Eschrig,Jiang}. As Eq.~(\ref{pot}) clearly shows, it can also affect the non-collinear spin ground state through the off-diagonal spin potential. Since the choice of on-site density has always some degree of ambiguity, it is important to carefully check this dependence. In the previous subsections, we compared two different local projectors, previously named as `full' and `dual' projector \cite{MJH}. The former takes the smaller $d$ occupation compared to the latter by construction \cite{MJH}. The calculation result of LNPO shows that the $J$ dependence of the canting angle is weaker when the `full' projector is used in both sFLL and cFLL. It is found that within sFLL $\theta = 1.88^\circ$ (`dual') and $2.20^\circ$ (`full') at $J=0$ eV which is in reasonable agreement with $\theta = 1.6^\circ$ obtained by using PAW (projector-augmented-wave) formalism \cite{Spaldin}. In cFLL, the difference between the two projectors is more pronounced as the off-diagonal potential is solely originated from interaction terms. In particular, for large $J > 0.7$ eV, the `dual' projector gives the negative $\theta$ whereas the `full' still favors the positive $\theta$. It is noted that such technical detail of choosing local projector can lead to a qualitatively different magnetic solution.

The $U$ dependence of canting angle is weaker than the case of $J$ for LNPO as reported in Ref.~\cite{Spaldin} while it was not negligible in SIO. We could not find a meaningful change in DM within our $J$ range. Instead, in order to understand this dependence, we once again investigate the explicit form of $\langle m_1,m_3|V_{ee}|m_4,m_2 \rangle$ in Eq.~(\ref{pot}), which is given in general by
\begin{eqnarray} \label{coulomb}
\langle m_1,m_2|V_{ee}|m_2,m_1 \rangle = \nonumber \\
\left(
\begin{array}{ccccc}
U+1.14J & 0.89J & 0.89J & 0.54J & 0.54J \\ 0.89J & U+1.14J & 0.43J & 0.77J & 0.77J \\ 0.89J & 0.43J & U+1.14J & 0.77J & 0.77J \\ 0.54J & 0.77J & 0.77J & U+1.14J & 0.77J \\ 0.54J & 0.77J & 0.77J & 0.77J & U+1.14J 
\end{array}
\right),
\end{eqnarray}
where the diagonalized DM is assumed with real harmonics basis  ($|d_{z^2}\rangle$, $|d_{x^2-y^2}\rangle$, $|d_{xy}\rangle$, $|d_{xz}\rangle$, and $|d_{yz}\rangle$). The conventional Slater parameterizations of $F^0=U$, $(F^2+F^4)/14=J$, and $F^4/F^2=0.625$ were used to generate Eq.~(\ref{coulomb}) \cite{Liechtenstein,Czyzyk,Vaugier}. One can notice that $U$ is included only in the diagonal part of Eq.~(\ref{coulomb}). As for the case of LNPO, the weak $U$ dependence of spin canting angle implies the dominant contribution of the inter-orbital exchange interactions (i.e., the off-diagonal elements of Eq.~(\ref{coulomb})) in determining the non-collinear ground state. Note that changing $F^4/F^2$ does not affect the diagonal elements of interaction matrix (Eq.~(\ref{coulomb})). Thus the canting angle change caused by different $F^4/F^2$ ratio shown in Fig.~\ref{fig3}(c) is also indicative of the role of inter-orbital exchange interactions. On the other hand, in the case of SIO where both $U$ and $J$ dependences are noticeable, the situation becomes more complicated, and not well understood within this simple formal analysis. As discussed within the collinear SDFT+$U$ case \cite{Mellan}, the $J$ dependence of Coulomb interaction matrix (Eq.~(\ref{coulomb})) gives rise to the nontrivial effect on the electronic structure. Our results of LNPO and SIO demonstrate that this is also the case for non-collinear magnetism.

\section{Concluding Remarks}

We performed a comparative study of DFT+$U$ formalisms for the case of non-collinear magnetic orders by examining the results of LNPO and SIO. The two formulations of cFLL and sFLL are found to give different $J$ dependences for the ground state magnetic order, and the difference is attributed to the spin-density XC functional which largely affects the canting angle especially through spin-off-diagonal potentials. Our result demonstrates that the inter-orbital exchange energy is a crucial component for non-collinear simulation and the care needs to be paid especially when SDFT+$U$ functional is adopted. Both sFLL and cFLL fail to reproduce the experimental canting angle in the reasonable range of interaction parameters for LNPO while cFLL gives a fairly good description for SIO. Our result poses a challenging issue for the capability and the accuracy of current DFT+$U$ and the related methods to describe the non-collinear magnetism in real materials. One promising aspect is that both formalisms give the qualitatively consistent results of $C_zA_x$ spin order for LNPO with experiments. Since our investigation is based on the fixed experimental lattice parameters, the discrepancy found in this study might become less significant when the lattice parameters are optimized. Further investigations at a more fundamental level are requested for the better first-principles description of non-collinear magnetic materials.

\section{Acknowledgements}
This work was supported by Basic Science Research Program through the National Research Foundation of Korea (NRF) funded by the Ministry of Education (2018R1A2B2005204). 

\vspace{0.5in}

\section*{References}

\bibliography{ref}

\end{document}